\documentclass[12pt]{jpconf}

\usepackage{graphicx}

\usepackage{subfig}

\begin{document}
\nocite{*}

\title{Oscillations in the bispectrum}

\author{P. Daniel Meerburg}

\address{Astronomical Institute ``Anton Pannekoek", University of Amsterdam, Amsterdam 1098 SJ, The Netherlands}

\ead{p.d.meerburg@uva.nl}

\begin{abstract}
There exist several models of inflation that produce primordial bispectra that contain a large number of oscillations. In this paper we discuss these models, and aim at finding a method of detecting such bispectra in the data. We explain how the recently proposed method of mode expansion of bispectra might be able to reconstruct these spectra from separable basis functions. Extracting these basis functions from the data might then lead to observational constraints on these models. 
\end{abstract}

\section{Introduction}
The aim of this review is to summarize the work done in \cite{Meerburg2010}, as presented at PASCOS 2010, Valencia in July of this year. In this work we investigate the method of mode expansion \cite{Fergusson,Fergusson:2009nv} for primordial bispectra that contain (a large number of) oscillations.  The discussed inflationary models/secanrios include features in the inflaton potential \cite{Chen:2006xjb,Chen:2008wn}, resonant models \cite{Chen:2008wn,Flauger:2009ab,Flauger:2010ja}, and models with a slight modification of the initial vacuum state \cite{Meerburg2009a,Meerburg2009b}. 

Although these models generally predict quite large primordial bispectral amplitude (i.e. $f_{NL}$), they are hard to constrain because their shapes are highly oscillatory (typically they also predict a large number of oscillations on top of the baryonic acoustic collations in the power spectrum, see e.g. \cite{Flauger:2009ab, Greene:2005aj}) A way to inquire unconstrained types of bispectra is to compare them with constrained types, such as local \cite{Gangui:1993tt, Komatsu:2001rj}, equilateral \cite{Creminelli:2005hu} and orthogonal \cite{Senatore:2009gt} type bispectra. The comparison can be qualified using the correlation function between two shapes in comoving momentum space defined as \cite{ Fergusson,Babich:2004gb}
\begin{eqnarray}
F_{X}\star F_{Y} & \equiv & \int_{\Delta_{k}}dk_{1}dk_{2}dk_{3} w_k S_{X}S_{Y}, 
\label{eq:dotproduct}
\end{eqnarray}
where $S$ is the shape of the bispectrum and $w_k$ is a weight function, which was chosen as $w_k=1/k_t$ in \cite{Fergusson:2009nv} to increase resemblance with the Fisher matrix (correlation) found in multipole space. The integral runs over the tetrahedral domain which is set by the triangle constraints together with the maximal observable $k_{max}$: 
\begin{eqnarray}
k_a\leq k_b+k_c\;\mathrm{for}\;k_a\geq k_b,k_c\nonumber\\
k_a,k_b,k_c\leq k_{max}\nonumber,
\end{eqnarray}
where $a,b,c=\{1,2,3\}$, $a\neq b\neq c$. Note that for optimal estimates one should in fact consider the correlation function in  multipole space, however this is computationally challenging and as a first estimate the above correlator is time efficient and quite precise \cite{Fergusson:2009nv}. 

Most constrained bispectra have a thing in common: they are relatively smooth. Consequently when comparing these to highly oscillating bispectra one finds that resulting constraints become relatively weak. In order to improve the analysis of these bispectra and the resulting bounds on associated physical parameters, one should directly constrain these models in the data. Unfortunately constraining bispectra from the data is very time consuming and computationally challanging, in particular since these models have two additional free parameters beside their amplitude: the phase ($\gamma$) and the frequency ($\omega_x$). In order to search for signatures in the data, one should scan through a range of possible frequencies and phases since it is possible to  completely miss the signal if one searches only for one specific/fixed frequency and phase. This compromises the analysis for these types of spectra to the extend that one should to try to consider other means.  

In a recent paper \cite{Fergusson,Fergusson:2009nv}, the authors propose a very elegant method of constraining any type of bispectra directly from  the data. The idea is to expand a given (theoretical) bispectrum into a set of orthogonal basis functions $R_n$, i.e. 
\begin{eqnarray}
S(x_1,x_2,x_3)&\simeq& \sum_{n=0}^N \alpha_n R_n(x_1,x_2,x_3).
\end{eqnarray}
where $x_i = k_i/k_{max}$, with $k_{max}$ the largest observable $k$ in the CMB, while $S(x_1,x_2,x_3)$, known as the (model-dependent) shape of the bispectrum, is defined via the primordial three point correlation function:
\begin{eqnarray}
\langle \zeta_{\vec{k}_1}  \zeta_{\vec{k}_2}  \zeta_{\vec{k}_3} \rangle&=& (2 \pi)^7 f_{NL}\Delta^2 \delta^K\left(\sum_{i=1}^3 k_i\right)\frac{S(k_1,k_2,k_3)}{k_1^2 k_2^2 k_3^2}.
\end{eqnarray}
Here $\zeta$ is the gauge invariant curvature perturbation ($\zeta = -H\delta \phi /\dot{\phi}_0$) which is constant after horizon exit, $\Delta$ is the amplitude of the primordial power spectrum (i.e. for single field slow-roll $\Delta=H^2/8\pi \epsilon$, where $H$ is the Hubble rate at the end of inflation and $\epsilon$ the slow-roll parameter). $f_{NL}$ is the amplitude of the primordial bispectrum and its value depend on the model under consideration. If the set of $R_n$ is optimally constructed and $S$ is relatively smooth, the number of modes necessary to reconstruct the spectrum is low. For example to reconstruct the bispectrum predicted by Dirac-Born-Infeld models of inflation (equilateral type) only $n=5$ modes lead to a 99\% correlation between the original and the reconstructed shape (in $k$ space) \cite{Meerburg2010, Fergusson:2009nv}. The advantage of this method of direct measurement is obvious. The constructed mode functions are aimed at being  able to reconstruct a large number of bispectra. Measuring these modes in the data could therefore constrain a large number of models at once (see \cite{Fergusson:2010dm}). As such, mode expansion can be considered an optimal way of extracting the bispectrum from the data, without focussing on specific shape (e.g. local and equilateral).

In the following (sec. \ref{middle_section}) we will use the method of mode expansion to a class of primordial bispectra that oscillate rapidly. We compare the existing set of mode functions, based on polynomials to a new set of base functions, based on Fourier functions. We show that the latter basis is more efficient in expanding oscillatory type bispectra. We will discuss (sec. \ref{Discussion}) some other related observations concerning the expansion, and explain how these could be useful in future data analysis. We will conclude in section \ref{Conclusion}.

\section{Models and their expansion}\label{middle_section}

In \cite{Meerburg2010} we investigated the ability to reconstruct inflationary bispectra that are not smooth using $R_n$ bispectral basis functions. We investigated three different shapes predicted by different physics:

\begin{eqnarray}
S_{Feat}&=& k_{max}^{-6}\sin (\omega_{f} x_t +\gamma_1), \label{eq:feat_bispectrum}
\end{eqnarray}
\begin{eqnarray}
S_{Res}&=& k_{max}^{-6} \sin (\omega_{r} \ln x_t+\gamma_2), \label{eq:res_bispectrum}
\end{eqnarray}
\begin{eqnarray}
S_{nBD} &=& \frac{\omega_{v}k_{max}^{-6}}{ x_1 x_2 x_3} \sum _j\frac{1}{x_j^3}\left(\frac{1}{2}\frac{\cos \left(\omega_{v} \frac{x_{j+1}+x_{j+2}}{x_{j}}+\gamma_3\right)}{\omega_v \left(\frac{x_{j+1}+x_{j+2}}{x_j}-1\right)} \right.- \frac{\sin \omega_{v} \left(\omega_{v} \frac{x_{j+1}+x_{j+2}}{x_{j}}+\gamma_3\right)}{\omega_v^2 \left(\frac{x_{j+1}+x_{j+2}}{x_j}-1\right)^2}\nonumber\\
&&\left. \frac{\cos \delta-  \cos \left(\omega_{v} \frac{x_{j+1}+x_{j+2}}{x_{j}}+\gamma_3\right) }{\omega_v^3 \left(\frac{x_{j+1}+x_{j+2}}{x_j}-1\right)^3}\right).
\label{eq:nbd_bispectrum}
\end{eqnarray}
For details explaining how to compute these bispectra and the non-Gaussian amplitude $f_{NL}$ for each of these models we would like to refer to \cite{Meerburg2010, Chen:2006xjb, Chen:2008wn, Flauger:2009ab, Flauger:2010ja, Meerburg2009a}. Generally, the various frequencies $\omega$ are model dependent as is the phase $\gamma$. Typically $\omega\gg\mathcal{O}(1)$. 

\begin{figure}[t]
   \centering
   \includegraphics[scale=.56]{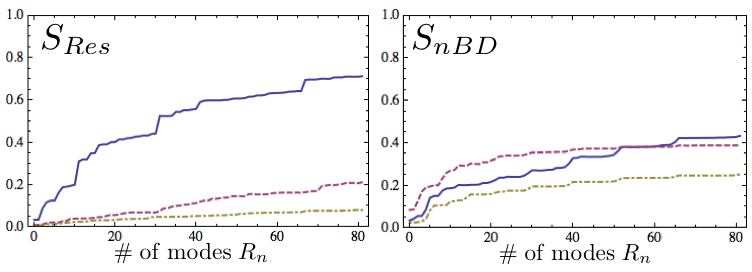} 
   \caption{The correlation of $S_{Res}$ and $S_{nBD}$ with their respective reconstruction using $R_n$ basis functions. The correlation is shown for three different frequencies: $\omega_{r,v}=20$ (solid), $40$ (dashed) and $60$ (dot-dashed).} 
    \label{fig:cosine_feat_res}
\end{figure}

As a first attempt we take the $R_n$ basis functions generated to be orthonormal on the tetrahedral domain. The details of $R_n$ mode construction on the this domain can be found in \cite{Fergusson:2009nv,Fergusson:2010dm}.
As expected, it requires many modes to establish significant correlation between the reconstructed shape $S^r$ and the predicted shape $S^p$.  In figure \ref{fig:cosine_feat_res} we show the various reconstruction attempts of eq. (\ref{eq:res_bispectrum}) and eq. (\ref{eq:nbd_bispectrum}) for several frequencies. 

\begin{figure}
  \centering
  \subfloat[Mode expansion for resonant type non-Gaussianity (eq. (\protect\ref{eq:res_bispectrum})) using the $\mathcal{F}_n$ basis functions. We considered the following frequecnies:  frequencies $\omega_r=20$ (blue, solid), $40$ (purple, dashed), $60$ (yellow, dot-dashed) and $80$ (green, solid). ]{\label{fig:expansion_a}\includegraphics[width=0.45\textwidth]{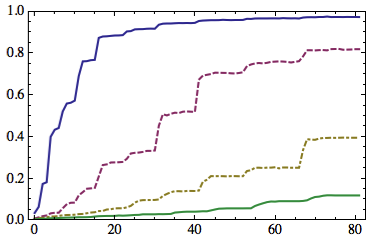}} \qquad
  \subfloat[Mode expansion for $S_{nBD}$ (eq.(\protect\ref{eq:nbd_bispectrum})) using the same mode functions as in \ref{fig:expansion_a}. Here we only considered two frequencies: $\omega_v=20$ (blue, solid) and $40$ (purple, dashed). The correlation seems to saturate at around 20\%. ]{\label{fig:expansion_b}\includegraphics[width=0.45\textwidth]{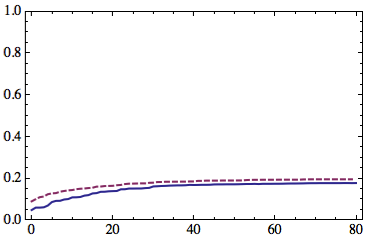}}\\
  \subfloat[Mode expansion comparison between $\mathcal{F}_n$ (top) and $R_n$ (bottom) with a fixed frequency for $S_1$. ]{\label{fig:expansion_c}\includegraphics[width=0.45\textwidth]{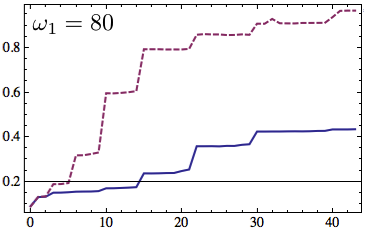}}\qquad
  \subfloat[Mode expansion comparison between $\mathcal{F}_n$ (top) and $R_n$ (bottom) with a fixed frequency for $S_2$.]{\label{fig:expansion_d}\includegraphics[width=0.45\textwidth]{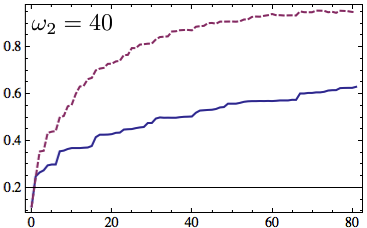}}
  \caption{Mode expansion for various models.}
  \label{fig:correlation_expansion}
\end{figure}

The $R_n$ basis functions are based on polynomials, e.g. $R_3 \propto x_1^3 +x_2^3 +x_3^3+c$ and higher $n$ correspond to larger powers of $x_i$. Such basis functions are not particularly optimized for oscillatory bispectra. In \cite{Meerburg2010} we considered another set of mode functions, denoted $\mathcal{F}_n$, based on $e^{-i\omega x}$ (Fourier reconstruction). It must be noted that $S_{Feat}$ is already of this form and therefore we did not consider this model in the expansion. To further investigate the reconstructive power of $\mathcal{F}_n$ we considered three additional toy model shapes:
\begin{eqnarray}
S_1&=&\left(\sin \frac{\omega_1}{x_1+1}+\sin \frac{\omega_1}{x_2+1}+\sin\frac{\omega_1}{x_3+1}\right),\\
S_2&=& \sin \omega_2 x_1 x_2 x_3,\\
S_3&=&\left(\sin \frac{\omega_3 x_t}{x_1+1}+\sin \frac{\omega_3 x_t}{x_2+1}+\sin\frac{\omega_3x_t}{x_3+1}\right).
\label{eq:toy_model}
\end{eqnarray}
We have plotted the number of modes versus the total correlation between the original model and the expansion for $S_{Res}$, $S_{nBD}$, $S_1$ and $S_2$ in figure \ref{fig:correlation_expansion}. As can be seen, for most of these bispectra reconstruction using $\mathcal{F}_n$ versus $R_n$  is much more effective, reducing the number of modes to reach similar correlation by a factor of 5. However, for n$S_{nBD}$, $\mathcal{F}_n$ reconstruction is not improved. The possible explanation why Fourier expansion is even worse than polynomial expansion for this type of bispectrum, seems to be related to the rapid change in frequency in a fixed direction. Fourier expansion is optimized for scale invariant frequencies. The polynomial expansion is simply optimized in reproducing as many different shapes as possible, explaining the observation that it is able to slowly increase correlation with the addition of modes while Fourier expansion seems to saturate around 20\%. Given the large enhancement of the amplitude $f_{NL}^{nBD}$ of this type of non-Gaussianity (which scales as $\omega_v^3$, see e.g. \cite{Meerburg2009a}, one might still be able to extract some information from that data even with such small correlations \cite{Meerburg2010b}.  
\section{Discussion}\label{Discussion}

In general the expansion of the oscillatory primordial bispectra becomes unavailing for really large frequencies, both using $\mathcal{F}_n$ and $R_n$. There are however a number of interesting observations which could make constraining and expanding oscillating bispectra much more viable. First of all the expansion in mode functions of the resonant bispectrum has a very discrete character; if you consider fig. \ref{fig:expansion_a} and fig \ref{fig:expansion_c}  only few modes actually contribute significantly to the convergence of the correlation. One could only try to expand the spectrum only in those modes, which could significantly reduce the number of modes necessary. Since the important modes seem to be related to the direction of propagation of the oscillation, we find that this conclusion is independent of the phase. In other words, only the value of the weights ($\alpha_n$) will differ, not the mode numbers ($n$) that are relevant for the expansion, e.g. the resonant shape $S_{Res}$ appears to propegate in the $k_t$ direction. As such only those mode numbers have a non-zero $\alpha$ that corresponds to base mode functions that contain $\mathcal{F}_n =\{1,e^{i n k_t}\}$.  

Secondly, given the discreteness of the correlation it is (obviously) not necessary to constrain all mode functions in the CMB data to get an indication of there is an oscillatory three point signal and what the possible frequency of this signal might be.  For resonant non-Gaussianities we only need to consider those modes that have a significantly large $\alpha$, and the measured value of the weights $\alpha$ would be a direct measure of the frequency. If one could extract the multipole projected Fourier modes that are responsible for most of the weight, this could in principle provide signatures of primordial bispectra with frequencies much larger than $\omega_r=80$ shown in fig. \ref{fig:expansion_a}. Measuring modes up to e.g. $n=100$ would not only provide information about the frequency of the signal, but could also hint on the type of primordial bispectrum. 

\section{Conclusion}\label{Conclusion}
We have investigated the viability of mode expansion for bispectra that contain oscillations. The motivation for investigating such features and their mode expansion, is that recently it has been shown that several scenarios or mechanisms can produce such features not only in the power spectrum, but also in the bispectrum. The appearance of oscillations in the bispectrum makes comparison with existing bispectral constraints, based on smooth bispectra, very inefficient and there exists substantial room for improvement. Polynomial expansion has been proposed to achieve factorization of a given theoretical bispectrum and we have investigated this for three different models. As expected, the larger the frequency of the primordial bispectrum, the more modes it requires to establish a reasonable approximation of the original spectrum. Fortunately, both the resonant and non-BD bispectrum have an amplitudes that scale with their frequency. Therefore, a small improvement in correlation could lead to a significant improvement in the ability to constrain the model by measuring these modes in the data and reconstructing the primordial signal.

Complementarily, we have proposed a different basis expansion, based on Fourier functions instead of polynomials. Such expansion is more relevant for resonant and non-BD scenario, since the feature bispectrum can already be transformed into Fourier modes analytically, using identities. We have shown that Fourier modes are much more efficient for the resonant bispectrum, reducing the number of modes necessary to establish the same correlation as polynomial modes by at least a factor of 5.  For the non-BD bispectrum both Fourier expansion and polynomial expansion are difficult. In the case of Fourier expansion correlation increases fast with the addition of modes, but quickly saturates to maximum of about 20\%. We believe that this is due to the exact form of the bispectrum, which has many small features near the edges of the tetrahedral domain. One might hope that some of these very small features are washed out when you compute the multipole equivalent. We hope to investigate this in a future attempt.  In addition we have investigated three toy-spectra, not based on any particular model, which have a different oscillating orientation compared to the three theoretical models. Expanding these in Fourier modes show similar improvement (only two have been shown) compared to polynomial expansion as the resonant bispectrum. In general, we therefore belief that Fourier expansion is much more effective in the expansion of oscillatory spectra compared to polynomial basis expansion.

We showed that for resonant inflation only a limited number of modes contribute significantly in reproducing the original bispectrum. This would allow to consider (observationally) only those modes that contribute substantially. This holds independent of the phase and frequency of the signal and is due to the specific form of this bispectrum, which oscillates (primarily) in the $k_t$ direction. Because the modes that are important for the reconstruction of the original bispectrum are independent of the frequency, this also implies that when one would observe these modes in the data one could in fact find evidence for much larger frequencies than discussed here, simply because for larger frequencies these modes will also matter but their respective weight ($\alpha$) will be smaller. Despite the fact that we could not optimally expand the non-BD bispectrum using Fourier modes, we did look into the three toy-sepctra. We found that other modes are important. Moreover, the modes that are important directly represent the orientation of the oscillating spectrum and could therefore discriminate between different bispectra quite effectively.  If this conclusion holds after forward projection into multipole space, measuring a number of Fourier mode functions in the CMB data would present an efficient way of deducing whether oscillations are present in the data and could give both an indication of the frequency and the shape of the primordial bispectrum. 

\ack
He would also like to thank the organizers for allowing to present this recent work. The author was supported by the Netherlands Organization for Scientific Research (NWO), NWO-toptalent grant 021.001.040. The author received additional fundings for attending this conference from the Leids Kerkhoven-Bosscha Foundation (LKBF).

\end{document}